\documentclass[twocolumn]{article}
\usepackage[final]{graphics}

\font\tenmsa=msam10
\font\sevenmsa=msam7
\font\fivemsa=msam5
\font\tenmsb=msbm10
\font\sevenmsb=msbm7
\font\fivemsb=msbm5
\newfam\msafam
\newfam\msbfam
\textfont\msafam=\tenmsa  \scriptfont\msafam=\sevenmsa
\scriptscriptfont\msafam=\fivemsa
\textfont\msbfam=\tenmsb  \scriptfont\msbfam=\sevenmsb
\scriptscriptfont\msbfam=\fivemsb

\global\mathchardef\lesssim "142E

\newcommand{\slL}{\raise.15ex\hbox{$/$}\kern-.53em\hbox{$L$}}
\newcommand{\slP}{\raise.15ex\hbox{$/$}\kern-.53em\hbox{$P$}}
\newcommand{\slR}{\raise.15ex\hbox{$/$}\kern-.53em\hbox{$R$}}
\newcommand{\slQ}{\raise.15ex\hbox{$/$}\kern-.53em\hbox{$Q$}}
\newcommand{\slK}{\raise.15ex\hbox{$/$}\kern-.53em\hbox{$K$}}
\newcommand{\slSigma}{\raise.15ex\hbox{$/$}\kern-.53em\hbox{$\Sigma$}}
\newcommand{\slcalP}{\raise.15ex\hbox{$/$}\kern-.63em\hbox{$\cal P$}}

\newcommand{\be}{\begin{equation}}
\newcommand{\ee}{\end{equation}}     
\newcommand{\bea}{\begin{eqnarray}}
\newcommand{\ena}{\end{eqnarray}}

\def\build#1\over#2{\mathrel{\mathop{\kern 0pt#1}\limits_{#2}}}

\font\tenimbf=cmmib10 at 10pt
\font\sevenimbf=cmmib10 at 7pt
\font\fiveimbf=cmmib10 at 5pt
\newfam\imbf
\textfont\imbf=\tenimbf
\scriptfont\imbf=\sevenimbf
\scriptscriptfont\imbf=\fiveimbf
\def\imb{\fam\imbf\tenimbf}

\begin{document}
\title{\bf{A simple criterion for effects\\
 beyond hard thermal loops}}
\author{F.~Gelis\\
Brookhaven National Laboratory,\\
Physics Department, Nuclear Theory,\\
Upton, NY-11973, USA}
\maketitle

\begin{abstract}
  Thermal amplitudes with ultrasoft momenta, which are not accessible
  by standard methods of perturbation theory, have recently attracted
  a lot of interest. However, the comparison of external momenta with
  the ultrasoft scale $g^2T\ln(1/g)$ is a too crude criterion, since
  amplitudes with hard external momenta can also be non perturbative,
  if these momenta are close enough to the light-cone.  In this
  letter, I give a more refined criterion to decide if an amplitude is
  non perturbative, that applies to all situations. In physical terms,
  this condition states that non-perturbative effects appear if the
  particles running in loops have to travel distances larger than
  their mean free path.
\end{abstract} 
\vskip 4mm 
\centerline{\hfill BNL-NT-00/17}

\section{Introduction}
Early attempts to calculate the damping rate of a gluon at rest by
using the bare perturbative expansion of thermal QCD led to
inconsistent results. It was realized by Braaten and
Pisarski \cite{HTL1}, and by Frenkel and Taylor
\cite{HTL2} that the resummation of 1-loop corrections
known as ``hard thermal loops'' (HTL in the following) is necessary
because they are of the same order of magnitude as their tree-level
counterparts for soft (momentum of order $gT$) external momenta. This
leads to an effective perturbative expansion \cite{HTL-eff}
for soft modes, in which thermal effects appear via modifications of
the propagators, and through non-local couplings. The physical origin
of this non-locality is easy to understand: soft modes can couple to a
hard (momentum of order $T$) thermalized parton of the plasma at
different points along its trajectory.

In the HTL framework, the hard modes are massless and propagate freely
in the plasma. Their masslessness has already been identified as a
source of collinear singularities when one has to deal with amplitudes
having external momenta close to the light-cone, as exemplified by the
calculation of the production rate of real photons
\cite{Photons-soft}. Flechsig and Rebhan \cite{FlechR1}
have shown that it is possible to slightly modify the definition of
hard thermal loops by giving a thermal mass to the particle running in
the loop, without altering the properties of HTLs.

HTLs can also be derived from kinetic theory \cite{Kinetic},
where they appear via a Vlasov (collisionless) equation. In this
framework, it was found that collisions become important at the
ultra-soft scale $g^2T\ln(1/g)$
\cite{Ultrasoft}.  In more physical terms,
such modes couple to very long wavelength density fluctuations of the
hard modes, and the propagation of hard modes over long distances is
affected by collisions.

However, there are examples of problems in which one has to take into
account collisions even if the external momentum scale is hard. Among
such problems is the production rate of hard photons, which is
sensitive to the collisions of hard quarks in the plasma
\cite{AurenGZ2}, a difficulty very similar to the one encountered for
ultra-soft amplitudes. This indicates that the condition $Q\lesssim
g^2T\ln(1/g)$ on the external momentum used in order to determine if
collisions are important is too crude to apply to situations like the
hard real photon production.

In this letter, I derive a more accurate criterion to decide if
effects beyond the HTLs (like collisions) are important in a given
problem. This condition reads $\lambda_{\rm mean}\lesssim\lambda_{\rm
  coh}(Q)$, where $\lambda_{\rm mean}$ is the mean free path of the
particle running in the loop, and $\lambda_{\rm coh}(Q)$ is a
coherence length constructed with the external momentum $Q$. This
condition reduces to $Q\lesssim g^2T\ln(1/g)$ only in special cases.

\section{Coherence length}
\begin{figure}[htbp]
\centerline{\resizebox*{!}{2.5cm}{\includegraphics{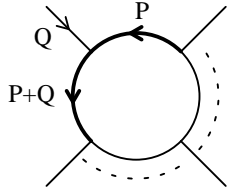}}}
\caption{Generic configuration of momenta. The boldface line 
  denotes the resummation of the width $\Gamma$ on the
  propagator.\label{fig:generic}}
\end{figure}

Let me consider a generic diagram evaluated at finite temperature in
which I isolate an arbitrary loop\footnote{This loop need not be a
  hard thermal loop.} having $Q$ as one of its external momenta.
Having in mind formalisms like the retarded-advanced formalism (or,
equivalently, the imaginary time formalism), we know that this loop is
evaluated by cutting each of the propagators of the loop in turn. Let
me focus particularly on the term where the propagator carrying the
momentum $P$ (see Fig.~\ref{fig:generic}) is cut.

The piece of interest to us is the product of the propagator carrying
the momentum $P$ and the propagator carrying the momentum $R\equiv
P+Q$ (i.e. the two propagators adjacent to the external line of
momentum $Q$):
\begin{equation}
[\Delta_{_{R}}(P)-\Delta_{_{A}}(P)]\Delta_{_{R}}(R)\; .
\end{equation}
The discontinuity in the square bracket is the cut propagator of
momentum $P$. The dominant term is obtained when we keep the product
of an advanced and a retarded propagator:
\begin{equation}
-\Delta_{_{A}}(P)\Delta_{_{R}}(R)\; .
\end{equation}
Indeed, it will turn out that these two propagators have very close
poles separated by the real energy axis. As a consequence the
integration contour cannot be moved to go around the poles, and this
configuration leads to large contributions.

For the sake of generality, we assume the propagators in the loop to
carry a mass $M$ and a collisional width $\Gamma\sim g^2T\ln(1/g)$.
Since the loop momentum is usually hard, the mass $M$ will typically
be the asymptotic mass $M_\infty\sim gT$ introduced by Flechsig and
Rebhan to improve the HTLs near the light-cone. The width $\Gamma$ is
a naive model for the collision term\footnote{Strictly speaking, the
  collision term of a Boltzmann equation does not necessarily lead to
  a complex pole in the propagator. It may happen that the propagator
  has a completely different analytical structure, without a pole
  \cite{Width}. Therefore, the width $\Gamma\sim
  g^2T\ln(1/g)$ can be seen as a model of the effect of collisions.
  This model is supported by the fact that the spectral function of
  the exact propagator is very similar to the spectral function of the
  propagator modeled by a width \cite{Width}.}  of a
Boltzmann equation, and sensitivity to this parameter in a loop
indicates that we are in a regime where collisions are important. In
this model, the inverse of the propagator reads
$\Delta_{_{R,A}}^{-1}(K)=(k_0\pm i\Gamma)^2-{\imb k}^2-M^2$.

Picking the value of $p_0$ at the pole of $\Delta_{_{A}}(P)$, and
plugging it in the denominator of $\Delta_{_{R}}(P+Q)$, we find:
\begin{equation}
\Delta_{_{R}}^{-1}(R)\approx 2\omega q_0-2pq\cos\theta
+Q^2+4i\Gamma (\omega+q_0)\; ,
\end{equation}
where we have neglected terms in $\Gamma^2$, and where $\theta$ is the
angle between the vectors $\imb p$ and $\imb q$. In the above
equation, $\omega$ denotes the real part of the pole of
$\Delta_{_{A}}(P)$, i.e. $\omega\equiv\pm\surd{({\imb p}^2+M^2)}$.

At this stage, we already see that the parameter $\Gamma$ can be
important even for large $Q$, provided that $q\approx q_0$, and
$\cos\theta\approx\pm 1$. In other words, the collisions manifest
themselves in perturbation theory via collinear singularities, and
$\Gamma$ acts as a collinear regulator. This regulator can be
extracted by setting $|\cos\theta|=1$.  Assuming also that $M\ll
p,\omega$, but keeping the external momentum $Q$ arbitrary, we obtain
\begin{equation}
\Delta_{_{R}}^{-1}(R)\approx 2\omega q \Big[1-{\rm sign}(\omega)\cos\theta
+{{M^2_{\rm eff}}\over{2\omega^2}}\Big]\; ,
\end{equation}
where we denote\footnote{If we were strictly in the HTL approximation,
  this quantity would be:
  \begin{equation}
    M^2_{\rm eff}{}_{|_{HTL}}=Q^2{{2\omega^2}\over{q_0(q_0+q)}}\; ,
    \end{equation}
    since at this level of approximation we have $M=0$, $\Gamma=0$,
    and we neglect $Q^2$ compared to $2P\cdot Q$.}:
\begin{equation}
M^2_{\rm eff}\equiv M^2+Q^2{{\omega(2\omega+q+q_0)}\over{q_0(q+q_0)}}
+4i\Gamma{{\omega (\omega+q_0)}\over{q_0}}\; .
\end{equation}
If $q_0\approx q$, this effective mass is identical to the one introduced in
\cite{AurenGZ2}.

We can see now that the width appears only in the imaginary part of
this effective mass. As a consequence, the condition to have an effect
due to collisions is simply ${\rm Re}\,M^2_{\rm eff}\lesssim {\rm
  Im}\,M^2_{\rm eff}$, or
\begin{equation}
(2\Gamma)^{-1}\lesssim \left[{{q_0 {\rm Re}\,M^2_{\rm eff}}
\over{2 \omega (\omega+q_0)}}\right]^{-1}\; .
\end{equation}
The left hand side of this inequality is nothing but the mean free
path $\lambda_{\rm mean}$ of the loop particle between two soft
scatterings. It is also possible to give a simple physical
interpretation to its right hand side. Indeed, the quantity in the
bracket gives the difference $\Delta E\equiv r^0-\omega_r$ between the
energy and its on-shell value for the particle of momentum $R$.  Its
inverse is the typical lifetime of this virtual state, and is usually
called coherence length and denoted by $\lambda_{\rm coh}(Q)$.  This
quantity has the physical interpretation of the length
traveled\footnote{By construction, the length scale obtained by
  setting $\cos\theta=\pm1$ is a longitudinal scale.} by the virtual
particle of momentum $P+Q$ before it fragments into an on-shell
particle of momentum $P$ and the external particle of momentum $Q$.
Therefore, the above inequality can be recasted in the more intuitive
condition:
\begin{equation}
\lambda_{\rm mean}\lesssim \lambda_{\rm coh}(Q)\; .
\label{eq:criterion}
\end{equation}
Eq.~(\ref{eq:criterion}) is the general condition for effects due to
collisions and its physical interpretation is the following: there is
a sensitivity to collisions if the loop particle travels distances
larger than the typical distance between two successive collisions. In
other words, $\lambda_{\rm coh}(Q)$ gives a quantitative measure of
``how much non-local'' is the effective coupling associated to the
loop.

One must also note that the above considerations must be repeated for
each external leg. There is a sensitivity to collisions if any of the
$\lambda_{\rm coh}(Q_i)$ one can define with the external momenta
$Q_i$ is larger than the mean free path.

As a preliminary check, we see that in the particular limit of almost
static fields ($q_0\ll q$), this condition becomes
$q\lesssim2\Gamma\sim g^2T\ln(1/g)$. In the opposite limit ($q\ll
q_0$) where the external particle is at rest , the condition similarly
becomes $q_0\lesssim2\Gamma\sim g^2T\ln(1/g)$. Therefore, in these two
limiting cases, Eq.~(\ref{eq:criterion}) is equivalent to the usual
condition $Q\lesssim g^2T\ln(1/g)$ used in \cite{Ultrasoft}.

\begin{figure}[htbp]
\centerline{\resizebox*{!}{4cm}{\includegraphics{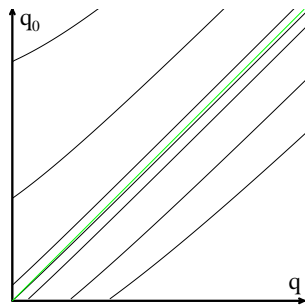}}}
\caption{Contour curves of $\lambda_{\rm coh}(Q)$.
  Large values of $\lambda_{\rm coh}(Q)$ are packed in the vicinity of
  the light-cone.\label{fig:phase-space}}
\end{figure}

However, the momentum $Q$ need not be ultra-soft in order for
Eq.~(\ref{eq:criterion}) to be satisfied. Indeed , it is trivial to
see that if $Q^2=0$ we have $\lambda_{\rm coh}^{-1}(Q)\approx
M^2/2\omega$ if $q_0\to+\infty$. Therefore, if $M\sim gT$, the
condition is satisfied even for arbitrarily hard external momenta,
provided they are on the light-cone. In order to make this discussion
more visual, Fig.~\ref{fig:phase-space} represents contour curves of
the quantity $\lambda_{\rm coh}(Q)$ in the $(q,q_0)$ plane.
From this plot, one can readily see that hard momenta close to the
light-cone are as ``dangerous'' as ultra-soft momenta.

In fact, the problem with hard momenta close to the light-cone could
also have been guessed from kinetic theory, where one must compare the
drift term $(v\cdot\partial_x)\delta N(k,x)$ and the collision term,
which in the relaxation time approximation can be written as $-\Gamma
\delta N(k,x)$. In momentum space, this comparison amounts to compare
$v\cdot Q$ with $\Gamma$, and one sees again the relative importance
of the collision term for large $Q$ if $Q^2\approx 0$.

\section{Vertex corrections}
Important corrections due to a collisional width on the loop
propagators are not the only manifestation of collisions in
perturbation theory. Some vertex corrections inside the loop also
become important when the condition of Eq.~(\ref{eq:criterion}) is
satisfied. More specifically, these vertex corrections are ladder
corrections connecting the two propagators adjacent to the external
line carrying the momentum $Q$.
\begin{figure}[htbp]
\centerline{\resizebox*{!}{2.5cm}{\includegraphics{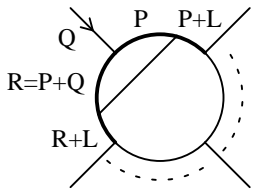}}}
\caption{Dominant corrections to the loop due
  to collisions. A boldface propagator is a propagator where a width
  has been resummed.\label{fig:corrections}}
\end{figure}
Indeed, one can easily see that each ladder correction increases the
degree of collinear divergence in the diagram, which compensates the
extra coupling constants. Each new rung in the ladder modifies the
result by a factor that can be estimated to be (see
Fig.~\ref{fig:corrections} for the notations):
\begin{equation}
I\equiv g^2 p r \int{{d^4L}\over{(2\pi)^4}}
\Delta_{_{A}}(P+L)\Delta_{_{R}}(R+L)n_{_{B}}(l_0) \rho(L)\; ,
\end{equation}
where $L$ is the momentum flowing in the rung, $\rho(L)$ is the
spectral function of the exchanged boson and $n_{_{B}}(l_0)$ its
statistical weight. Having in mind applications in QCD, we added the
factor $pr$ that gives the correct order of magnitude for the
numerator of quark propagators, or for the momentum dependence coming
from the 3-gluon coupling. In order to estimate the integral, it is
convenient to use the integration variables $l_0$, the transverse
(with respect to ${\imb q}$) momentum ${\imb l}_\perp$ and the
longitudinal momentum $l_z$. Performing the $l_z$ integral in the
complex plane by using the poles of $\Delta_{_{A}}(P+L)$, we can
estimate\footnote{Here, we do not consider the poles of the spectral
  function $\rho(L)$. This approximation, known as the ``approximation
  of independent scatterings'', is valid if the screening length
  $\mu^{-1}$ is smaller than the mean free path $\Gamma^{-1}$. This is
  not satisfied for transverse gluons, and the corresponding
  contribution has been studied in \cite{AurenGZ1}. It is important
  when the coherence length becomes larger than the screening length,
  and is not related to the collinear singularities we are studying
  here.}
\begin{equation}
I\propto g^2T {{pr}\over{q_0}} \int d^2{\imb l}_\perp {{dl_0}\over{l_0}} 
\rho(l_0,{\imb l}_\perp){1\over{({\imb p}_\perp+{\imb l}_\perp)^2+M^2_{\rm eff}}}\; ,
\end{equation}
where $M^2_{\rm eff}$ is the effective mass introduced earlier. The
value of $l_z$ at the pole of $\Delta_{_{A}}(P+L)$ being very small,
we neglect the $l_z$ dependence of the spectral function $\rho$.
Because the second denominator contains ${\imb p}_\perp+{\imb
  l}_\perp$, the second loop cannot be factorized from the first one.
However, the order of magnitude of ${\imb p}_\perp$ is also $M_{\rm
  eff}$ (because of a denominator ${\imb p}_\perp^2+M^2_{\rm eff}$ in
the first loop), so that for the sole purpose of estimating $I$ we can
just replace ${\imb p}_\perp+{\imb l}_\perp$ by ${\imb l}_\perp$. In
addition, one can use sum rules to perform the $l_0$ integral:
\begin{equation}
I\sim g^2T {{pr}\over{q_0}}\int d^2{\imb l}_\perp 
{1\over{{\imb l}_\perp^2+\mu^2}}{1\over{{\imb l}_\perp^2+M^2_{\rm eff}}}\; ,
\end{equation}
where $\mu$ is a Debye mass (of order $gT$) if the exchanged gauge
boson is longitudinal, or the magnetic mass (of order $g^2 T$) if this
boson is transverse. Up to some inessential factors, the above
integral is
\begin{equation}
I\sim g^2T \ln\Big({{M^2_{\rm eff}}\over{\mu^2}}\Big) {{pr}\over{q_0(M^2_{\rm eff}-\mu^2)}}\; .
\end{equation}
Note that we have always $\mu\lesssim M_{\rm eff}$. If we have $\mu\ll
M_{\rm eff}$ (this is what happens for a transverse gluon since then
$\mu\sim g^2T$), the above result simplifies into
\begin{equation}
I\sim g^2T \ln\Big({{M_{\rm eff}}\over{\mu}}\Big) 
{{pr}\over{q_0 M^2_{\rm eff}}}\sim \Gamma {{pr}\over{q_0 M^2_{\rm eff}}}\; .
\label{eq:ladder}
\end{equation}
If on the contrary we have $\mu\approx M_{\rm eff}$ (which can happen
for a longitudinal gluon), it becomes\footnote{We can still identify
  $\Gamma$ in the result, because if $\mu$ is large enough to have
  $\mu\approx M_{\rm eff}$, then the damping rate $\Gamma$ has an
  infrared cutoff large enough to suppress the logarithm.}
\begin{equation}
I\sim g^2T 
{{pr}\over{q_0 M^2_{\rm eff}}}\sim \Gamma {{pr}\over{q_0 M^2_{\rm eff}}}\; .
\label{eq:ladder1}
\end{equation}

We have now to consider two cases, depending on whether $M^2_{\rm
  eff}$ is dominated by its real part or imaginary part:
\begin{eqnarray}
&&{\rm If\ \ }\lambda_{\rm coh}(Q)\ll \lambda_{\rm mean}\;,
\quad I\sim {{\lambda_{\rm coh}(Q)}\over{\lambda_{\rm mean}}}\ll 1\; ,\nonumber\\
&&{\rm If\ \ }\lambda_{\rm mean}\lesssim \lambda_{\rm coh}(Q)\;,
\qquad I\sim{{\lambda_{\rm mean}}\over{\lambda_{\rm mean}}}=1\; .
\end{eqnarray}

Therefore, we see from this estimate that each new rung in the ladder
brings an extra contribution which is of order $1$ if
Eq.~(\ref{eq:criterion}) is satisfied (both for transverse and
longitudinal gluons, contrary to the term studied in \cite{AurenGZ1}),
indicating that ladder corrections must be resummed whenever the
effect of the width $\Gamma$ is important. It is worth noting that the
mechanism that makes ladder corrections important in this context is
related to collinear singularities (the variable ${\imb l}_\perp$ can
be related to an angular deviation), that do not distinguish
transverse and longitudinal gluons.

It may happen that the vertex corrections actually cancel the
resummation of $\Gamma$ \cite{Lebed-smilga}. This is not the case in
QCD, or for some specific problems like the photon
production
rate by a quark-gluon plasma \cite{AurenGZ2}.  When they occur, these
cancellations are the sign that the relevant mean free path is
actually larger than the inverse $\Gamma^{-1}$ of the damping rate
(usually $1/g^4T\ln(1/g)$ instead of $1/g^2T\ln(1/g)$)\footnote{From
  this perspective, there seems to be a difference between photon
  production and momentum transport for instance: the relevant
  scatterings for momentum transport are hard and their rate is small
  ($g^4T\ln(1/g)$), while photons (soft, or hard but collinear) can be
  produced in a quark gluon plasma via bremsstrahlung induced by soft
  scatterings of the quarks, which are more frequent (rate
  $g^2T\ln(1/g)$).}.  The simple discussion presented in this paper
cannot exclude such cancellations, and they must be studied on a case
by case basis. In the language of kinetic theory, these cancellations
appear immediately in the collision term of the Boltzmann equation.

As a side note, these vertex corrections were also to be expected on
the basis of Ward identities in gauge theories. The fact that we
obtain for them the same criterion as for the corrections by the width
can therefore be seen as a consistency check.

\section{Examples}
In this section, I list a few examples where use of
Eq.~(\ref{eq:criterion}) can tell immediately if the problem is
tractable by perturbative methods (i.e. without going beyond hard
thermal loops) or not.

\subsection{Hard Thermal Loops}
The criterion $\lambda_{\rm mean}\lesssim \lambda_{\rm coh}(Q)$ can be
applied to determine in which kinematical regime the hard thermal
loops themselves should be corrected by effects due to collisions. We
see that the situation $Q\lesssim g^2T\ln(1/g)$ studied in
\cite{Ultrasoft} is not the only domain where such
corrections are important. Indeed, HTLs should also be corrected for
soft momenta\footnote{The derivation of HTLs in \cite{HTL1} assumed
  generic soft external momenta, far enough from the light-cone.}
close to the light cone. From a technical perspective, this is due to
collinear singularities that show up in HTLs when they have external
legs close to the light-cone, and the asymptotic mass $M_\infty$
advocated in \cite{FlechR1} to solve this problem is not the most
relevant regulator.

\subsection{Viscosity}
The shear viscosity has been calculated in a scalar theory by Jeon in
\cite{Jeon2} and more recently in \cite{Viscosity}, and it was
noticed that this quantity receives contributions from an infinite
series of diagrams. In thermal field theory, this transport
coefficient is obtained as the imaginary part of the correlator of two
energy-momentum tensors, in the limit $q=0,q_0\to 0$. Therefore, this
corresponds to a point in figure \ref{fig:phase-space} where the coherence
length $\lambda_{\rm coh}(Q)$ is infinite, which explains why this
quantity is severely non-perturbative.

\subsection{Photon production}
In \cite{AurenGZ2}, the photon production rate of quasi-real photons
by a quark-gluon plasma has been found to be sensitive to multiple
scatterings undergone by the quark that emits the photon, even for a
hard photon. The criterion for this effect is precisely $\lambda_{\rm
  mean}\lesssim \lambda_{\rm coh}(Q)$ where $Q$ is the momentum of the
produced photon. In this particular situation, the effect of
collisions is known as the Landau-Pomeranchuk-Migdal effect, and the
coherence length $\lambda_{\rm coh}(Q)$ has also the interpretation of
the formation time of the photon.

\subsection{Damping rate of a fast fermion}
The perturbative calculation of the damping rate of a hard fermion
suffers also from collinear singularities
\cite{Damping}, which can be understood in the
present framework because the condition Eq.(\ref{eq:criterion}) is
satisfied for this problem (the fermion is hard, but on-shell so that
the corresponding $\lambda_{\rm coh}(Q)$ is large.). The
non-perturbative study of this problem has been performed by Blaizot
and Iancu in \cite{Width}, and led to a non-exponential
decay of the propagator for large time differences.

\subsection{Out-of-equilibrium systems}
As a side note, we can also mention another area where the coherence
length defined in this paper should prove useful: that of out of
equilibrium systems. New questions arise when the system is
out-of-equilibrium: does it make sense to define a local (in
space-time) production rate? can one ignore effects of the relaxation
towards equilibrium in effective couplings? To answer these questions,
one should compare the coherence length with the relaxation time of
the system, or with the typical length scale of spatial
inhomogeneities. Indeed, if the coherence length $\lambda_{\rm
  coh}(Q)$ associated with the external leg of some effective coupling
is larger than the relaxation time of the system, then this coupling
should receive corrections reflecting the fact that the system is out
of equilibrium. In imaged terms, the effective coupling is so
non-local that it becomes sensitive to the large scale of gradients in
the system.

\section{Conclusions}
In thermal field theory, effects from collisions manifest themselves
through collinear and/or infrared singularities that enhance otherwise
suppressed higher order diagrams.  By studying the behavior of some of
those diagrams, we have derived a condition under which collisions
provide important corrections to an amplitude having an external
momentum $Q$, that generalizes the usual $Q\lesssim g^2T\ln(1/g)$ used
in previous works.  This condition reads
\begin{equation}
\lambda_{\rm mean}\lesssim \lambda_{\rm coh}(Q)\; ,
\end{equation}
where $\lambda_{\rm mean}$ is the mean free path of the particle
running in the loop, and $\lambda_{\rm coh}(Q)$ is the typical length
this particle has to propagate. The interpretation of the criterion is
therefore straightforward: collisions are essential if the particle
running in the loop travels distances larger than the average distance
between two collisions.

This criterion applies to a wide range of problems that were known to
be non-perturbative, including problems involving hard momenta (that
would have been misleadingly classified as perturbative if one uses
the criterion $Q\lesssim g^2T\ln(1/g)$), highlighting common features
of seemingly unrelated problems.
\vskip 2mm
\noindent{\bf Acknowledgements:}
I would like to thank P.~Aurenche, A.~Peshier, R.~Pisarski, and
H.~Zaraket for useful discussions and suggestions.  This work is
supported by DOE under grant DE-AC02-98CH10886.

\end{document}